\documentclass[conference]{IEEEtran}
\IEEEoverridecommandlockouts
\usepackage{cite}
\usepackage{amsmath,amssymb,amsfonts}
\usepackage{algorithmic}
\usepackage{graphicx}
\usepackage{textcomp}
\usepackage{tcolorbox}
\usepackage{listings}
\usepackage{color}

\definecolor{dkgreen}{rgb}{0,0.6,0}
\definecolor{gray}{rgb}{0.5,0.5,0.5}
\definecolor{mauve}{rgb}{0.58,0,0.82}

\def\BibTeX{{\rm B\kern-.05em{\sc i\kern-.025em b}\kern-.08em
    T\kern-.1667em\lower.7ex\hbox{E}\kern-.125emX}}

\begin{document}

\title{Refactoring Software in the Automotive Domain \\for Execution on Heterogeneous Platforms}

\author{\IEEEauthorblockN{Hugo Andrade}
\IEEEauthorblockA{\textit{Computer Science and Engineering} \\
\textit{Chalmers University of Technology}\\
Gothenburg, Sweden \\
sica@chalmers.se}
\and
\IEEEauthorblockN{Ivica Crnkovic}
\IEEEauthorblockA{\textit{Computer Science and Engineering} \\
\textit{Chalmers University of Technology}\\
Gothenburg, Sweden \\
ivica.crnkovic@chalmers.se}
\and
\IEEEauthorblockN{Jan Bosch}
\IEEEauthorblockA{\textit{Computer Science and Engineering} \\
\textit{Chalmers University of Technology}\\
Gothenburg, Sweden \\
jan.bosch@chalmers.se}
}

\maketitle

\begin{abstract}
The most important way to achieve higher performance in computer systems is through heterogeneous computing, i.e., by adopting hardware platforms containing more than one type of processor, such as CPUs, GPUs, and FPGAs. Several types of algorithms can be executed significantly faster on a heterogeneous platform. However, migrating CPU-executable software to other types of execution platforms poses a number of challenges to software engineering. 
Significant efforts are required in such type of migration, particularly for re-architecting and re-implementing the software. Further, optimizing it in terms of performance and other runtime properties can be very challenging, making the process complex, expensive, and error-prone. Therefore, a systematic approach based on explicit and justified architectural decisions is needed for a successful refactoring process from a homogeneous to a heterogeneous platform. In this paper, we propose a decision framework that supports engineers when refactoring software systems to accommodate heterogeneous platforms. It includes the assessment of important factors in order to minimize the risk of recurrent problems in the process. Through a set of questions, practitioners are able to formulate answers that will help in making appropriate architectural decisions to accommodate heterogeneous platforms. The contents of the framework have been developed and evolved based on discussions with architects and developers in the automotive domain. 
\end{abstract}


\vspace{-0.2 cm}

\section{Introduction}
As technology advances and software applications become widespread, the requirements for multiple functionalities increase at a fast rate. In the automotive industry, high-end products nowadays embed more than 100 million lines of code that realize a variety of functions. From robust safety features to increased comfort in the cabin, the role of software has taken as much importance as the mechanical integrity of the vehicles. The industry now has a clear focus on artificial intelligence (AI) applications that handle very large amounts of data. Mainly due to such large amounts of data, most time and development efforts in this domain are spent on understanding, preparing, monitoring, and logging of data, rather than actually implementing the machine learning algorithms and models \cite{Bernardi2019}. Such high demands on software can only be realized through mechanisms that allow for increased hardware performance and energy efficiency at a reasonable cost. 

Currently, the most important way to increase performance of computer systems is by using heterogeneous platforms, i.e., hardware platforms containing more than one type of processor, like CPUs, GPUs, and FPGAs. In the context of heterogeneous platforms, the processing of data can be parallelized, and different types of data can be assigned to specialized processors. For instance, GPUs are known to be more efficient than CPUs when executing tasks that require multiple parallel processes. A typical example is computer vision, which processes image data obtained from sensors to create an accurate world model. Heterogeneous computing, however, poses a number of challenges of software engineering, mainly due to the inherently different characteristics of the hardware processing units. 
It is typically very difficult to optimize the processing of data with respect to non-functional requirements such as performance, energy consumption, and real-time constraints. 

In most industrial cases, new products are developed from existing software. The challenges are not only related to the deployment of software onto a heterogeneous platform, but also to the refactoring of existing software for it to be executed on a new platform. This scenario poses a number of new challenges particularly to the software architecture design -- which have not been addressed in literature, to the best of our knowledge. However, we have identified through our industrial partners a major need for a systematic approach to support decision-making during such migration process -- from CPU-centric to heterogeneous platforms.

In this paper, we propose a reasoning framework that specifies a set of considerations supporting the decision-making process in refactoring software systems when migrating from CPU-centric to heterogeneous platforms. We provide means for reasoning about different aspects that practitioners must address when refactoring software-intensive systems. Our proposal is based on a series of in-depth discussions with our industrial partners in the automotive industry. 


The remainder of this paper is organized as follows. 
A motivational example is presented in Section II.
In Section III, we present the research methodology that was used in this study. In Section IV, we describe our approach to refactoring systems for heterogeneous platforms.  We discuss validation of the proposed framework in Section V. In Section VI, we present the related work. Finally, in Section VII, we present the conclusion and future work.

\section{Motivational example}
\label{sec:motivation}
The automotive industry provides an illustrative example of architectural evolution of a system deploying heterogeneous computing.

A current state of practice for system and software architecture for vehicular control systems, working well in the last 25 years, is a distributed system consisting of many computational units (a.k.a. Electronic Control Units -- ECUs) with embedded software, typically including a control loop that receives signals from sensors, performs computation, and produces signals to the connected actuators that control the electromechanical parts of the vehicle. For the communication between ECUs, a common bus (typically a standard Controller Area Network (CAN)) is used. 
This modular and component-based approach, like AUTOSAR \cite{AUTOSAR}, enables efficient evolution of the system: to introduce a new service, a new ECU is added with its embedded software. ECUs use simple CPUs and are dimensioned to maximally utilize the computational and memory resources that are automotive grade.

In recent years, the development of new software and hardware technologies has enabled significant improvements in the automotive industry.
The main and disruptive changes are the transformation to electrical vehicles, autonomous driving, and connectivity. Examples of new functionality include different elements of autonomous driving, optimized engine control, and improved behavior in risk-full situations.

The new functions being introduced are typically computation intensive, with extensive parallel computing, processing large amounts of data in real-time, and have major requirements on system performance. The new technologies include the use of machine learning, parallel computing, intensive communication in real time, cloud computing and edge-computing, etc. This requires that many strategically important architectural decisions need to be made with respect to (i) system and software architecture; and (ii) business-oriented decisions on development and deployment processes.

\begin{figure}[b!]
\centering
\vspace{-0.4cm}

\includegraphics[scale=0.38]{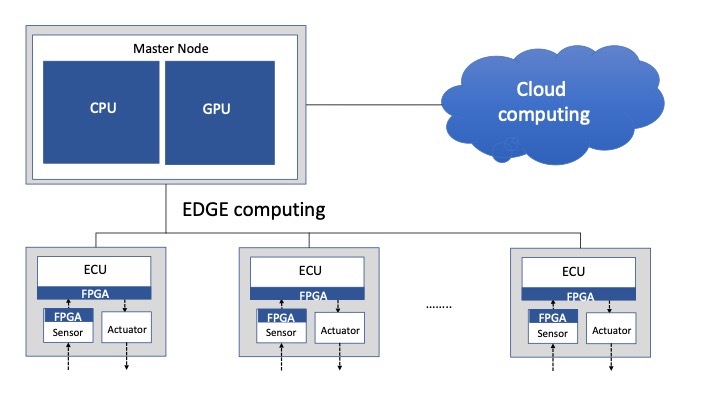}
\caption{Example of vehicular control system architecture}
\vspace{-0.2cm}
\label{fig:new-car-architecture}
\vspace{-0.2cm}
\end{figure}

Fig~\ref{fig:new-car-architecture} shows a new architecture of an automotive system that provides architectural prerequisites for new functions and enables a continuous transformation from the old architecture to the new architecture. 

The basic architecture is the same -- distributed systems connect via a bus, but the node structure is changing. Instead of nodes optimized for low computational and storage capacity, the nodes (ECUs) become heterogeneous computational platforms: (i) CPUs are getting more powerful, and in some cases replaced by multi-core CPUs, (ii) FPGAs are being included on the platforms for specialized computation, in particular processing input data from sensors (such as camera and radars). Further, the sensors are equipped with computational platforms (typically CPU + FPGA), enabling direct data processing and significant reduction in the amount of data for further processing. Additionally, the computing power is concentrated on a new, centralized, powerful computational platform that includes (multi-core) CPUs and GPUs, and in this way, many functions from distributed ECUs can be moved to it. Thus, the most-intensive computational services can be performed in real-time. This computation altogether can be seen as \textit{edge computing} in respect to the cloud computing to which the automotive system is connected, but not necessarily continuously. The cloud computing resources are used for additional services that do not have hard real-time requirements. Additionally, the cloud computing resources are used for further development of the system, including the training of machine learning models and analysis of the data provided by the monitoring and logging functions of the vehicles. 


Refactoring to heterogeneous platforms requires rewriting the code. The following code snippet illustrates the type of code changes that is required \cite{CUDAintro}. The example depicts a simple function in C++ using the CUDA framework \cite{CUDA} that adds the elements of two arrays. Compared to plain C++ code, the add function must be transformed into a function that can be executed on the GPU by adding the global specifier. Then, the memory must be explicitly allocated in a location that can be accessible by the GPU. In this example, the memory space is accessible by both the CPU and the GPU. Finally, with a number of changes to the syntax, the ``add'' function is invoked on the GPU using multiple parameters. CUDA programming demands explicit management of several aspects, such as device memory, data transfer between memories, and the synchronization of data access. For instance, the CPU must be told to wait for the GPU to finish the job before accessing the data in a shared memory.


\begin{lstlisting}[basicstyle=\footnotesize, float,floatplacement=H, caption={Example code written in C++ using the CUDA framework, showing the implications of the migration in terms of syntax and architecture decisions.},captionpos=b, belowskip=-0.8 \baselineskip ]
#include <iostream>
#include <math.h>
// Kernel function to add the elements of two arrays
__global__
void add(int n, float *x, float *y)
{
  for (int i = 0; i < n; i++)
    y[i] = x[i] + y[i];
}

int main(void)
{
  int N = 1<<20;
  float *x, *y;

  // Allocate Unified Memory -- accessible from CPU or GPU
  cudaMallocManaged(&x, N*sizeof(float));
  cudaMallocManaged(&y, N*sizeof(float));

  // initialize x and y arrays on the host
  for (int i = 0; i < N; i++) {
    x[i] = 1.0f; y[i] = 2.0f;
  }

  // Run kernel on 1M elements on the GPU
  add<<<1, 1>>>(N, x, y);

  // Wait for GPU to finish before accessing on host
  cudaDeviceSynchronize();

  // Check for errors (all values should be 3.0f)
  float maxError = 0.0f;
  for (int i = 0; i < N; i++)
    maxError = fmax(maxError, fabs(y[i]-3.0f));
  std::cout << "Max error: " << maxError << std::endl;

  // Free memory
  cudaFree(x); cudaFree(y);
  
  return 0;
}

\end{lstlisting}

\normalsize


The aforementioned example setting shows the complexity of the process, which raises important questions related to the process of refactoring software. The affected areas include evaluation, design, testing and deployment operations. Some of these challenges are listed next. 
\begin{itemize}
\item Process-related decisions:
\begin{itemize}
    \item What is the process of refactoring of the system architecture? 
    \item What is the process of refactoring of existing code from a platform (CPU) to a heterogeneous platform (e.g., GPU, or FPGA)?
    \item What are the implications of deployment of new architectures on the overall system's properties (resource utilization, overall system performance, development and production costs, etc.)?
\end{itemize}

\item Deployment of a new architecture:
\begin{itemize}
    \item How to transform the existing architecture into the new architecture? 
    \item How to distribute services in the new architecture, and ensure quality properties related to real-time requirements, performance, resource utilization, etc? 
\end{itemize}
\end{itemize}

These questions have many sub-questions and require many decisions of different types. Additionally, the decisions are interdependent. For this reason, it is very challenging to successfully manage software refactoring in such setting.
Based on the aforementioned topics, and considering the level of complexity involved in the process of refactoring, a systematic approach that defines the decision process and its implementation is needed.

\section{Research methodology}
\label{sec:researchmethodology}

This study was conducted with basis on the design science methodology \cite{Dresch2014}. In summary, design science research (or \textit{constructive science research}) aims to establish and operationalize research when the desired goal is an artifact or a recommendation. The procedures of design science research culminate in new ideas or a set of analytical techniques that enable the development of research \cite{Vaishnavi2007}. The proposed framework was created based on two different perspectives: the state-of-the-art and the experiences of practitioners in large industrial contexts. 

First, we studied the available literature on the topics of software deployment \cite{Andrade2019}
and software architectures \cite{Andrade2018} for heterogeneous platforms in order to obtain an overview of the research area and identify the challenges, concerns, and gaps in research. In this stage, we identified a number of approaches to realize software architecture when a heterogeneous platform is available. In summary, the architectural approaches can be organized into two main categories: architectural principles, and architectural styles (or patterns). These different techniques were studied and incorporated into our proposed approach as software architecture design alternatives (see \ref{sec:archdesign}).

Then, we iteratively identified the gaps between theory and practice in discussions with multiple partners in industry, capturing their practices, and analyzing their perspectives in trying to meet different challenges \cite{Andrade2019}. We sketched the first versions of the framework and held multiple discussions among the authors for adjustments. When it had reached a certain level of maturity, we scheduled face-to-face meetings with our industrial partners in order to present it. These meetings were conducted in the form of workshops, in which we presented the framework and gave the attendants the opportunity to discuss topics of interest. 

Finally, we validated the approach through a questionnaire that was sent out electronically to multiple companies. It was designed to capture the practitioners' impressions of the framework through explicit, open questions. The responses were qualitatively analyzed and the suggestions incorporated into the framework.

\section{Refactoring for heterogeneous platforms}
\label{sec:Re-architecting}

In this section we describe the framework, which consists of four steps that are explained in detail below. The steps are, namely: \textit{A. ``Determining the impact on the software architecture''}; \textit{B. ``Mapping software and hardware''}; \textit{C. ``Determining the overall architecture design''}; and \textit{D. ``Refactoring software components''}. Within each step we elaborated activities and questions that should be answered by the system engineers in order to obtain a set of considerations of different perspectives. Finally, the answers to these questions will help the engineers to make appropriate architectural decisions.

\subsection{Determining the impact on the software architecture} 
When introducing a new processing unit, the software architecture must be adapted to accommodate the changes. In particular, issues related to communication and memory management become relevant.

A1: \textit{Examine the existing data pipeline.}
As first step, engineers must examine the existing software architecture in order to obtain an understanding of the current design. In particular, the communication between components should be revisited as the data pipeline will be changed with the introduction of a heterogeneous platform. In this stage, a reassessment of the system's documentation might be useful, given that there is consistency between the documentation and the actual implementation. Engineers should run measurements to obtain a ``default’’ performance of the system prior to refactoring.

A2: \textit{Determine the expected performance gains.}
Then, the engineers specify which non-functional properties are intended to be improved, according to the system’s requirements. One should particularly take into account the additional communication demands that will be present in such a distributed system. In the case of automotive applications, there are substantial constraints in terms of resources that can be utilized. This issue is partially addressed if engineers have the liberty to first design software -- putting software functionalities in the focus -- and then proceed to determine the embedded hardware components that will be utilized. The assessment can include several considerations related to non-functional properties concerning runtime (e.g., timing issue, power consumption), life-cycle (e.g., development process, maintainability), or business models (e.g., development and production costs)\cite{Sapienza2014}.

A3: \textit{Elicit the changes in the software architecture.}
Engineers must then elicit the necessary changes in the software architecture according to the predefined non-functional requirements. The communication between components become relevant to the performance of the overall system, due to the inherent characteristic of heterogeneous systems to pass over messages to components deployed on accelerators. For instance, one decision in this stage may indicate that the messages required by computationally heavy functionalities will need to be forwarded to the newly introduced software component to be deployed on the heterogeneous platform. Therefore, the message passing infrastructure must ensure the adequate capabilities for such communication to occur.


\begin{tcolorbox}
\textbf{Determining the impact on the software architecture}


Q1: What is the current status of the data pipeline? 

Q2: What are the expected gains in performance?

Q3: What changes are necessary in the architecture?
\end{tcolorbox}

\vspace{-0.2cm}

\subsection{Mapping software and hardware}
Determining the mapping between software and hardware can be very challenging, typically requiring several rounds of experimentation and prototyping. 

B1: \textit{Identify the functionalities to be accelerated.} 
The accelerated portions of the software will most likely be the ones identified as the most computationally intensive tasks. In the case of automotive applications using AI technology, for instance, the training of the machine learning algorithms are strong candidates for execution on the accelerator(s). These algorithms typically include the processing of multi-dimensional matrices that are suitable for execution on GPUs.

B2: \textit{Experiment and measure performance.}
Well-established frameworks, such as CUDA, typically embed a useful tool for assessing portions of code in terms of execution time, namely ``profiling''. With this tool, practitioners can experiment with different portions of code prior to determining the mapping between software and hardware. This step allows engineers to assess the performance of potential configurations and compare them to the default benchmark obtained with the original CPU-centric software.

B3: \textit{Establish a configuration that is suitable.}
A number of approaches have been proposed in literature to tackle the mapping between software and heterogeneous platforms. In \cite{Svogor2019} for instance, the authors used a genetic algorithm to find a locally optimal solution in respect to a defined cost function. The model proposed in the paper takes into consideration both the system constraints and user-defined architectural decisions. The latter might include, for instance, a requirement that two particular components are not allowed to be allocated to the same processing unit. All constraints are accounted for in a cost function, representing the overall performance of the system given a certain allocation configuration. The result of the proposed method is a system deployment configuration that is (at least) nearly optimal for the overall system performance. 

Additionally, dynamic deployment mechanisms can be created in order to allow different components to be executed depending on the current status of runtime. There is plenty of literature that defines the mapping procedure \cite{Andrade2019} -- most often in selecting particular concerns and defining a cost function, attempting to find its minimum in respect to a given component distribution. There are some approaches that enable the specification of many requirements and/or resource constraints, as well as the communication capacity required for interaction between the components, attempting to find a local optimum for a set of components\cite{Svogor2019,Santana2015}. These approaches may be challenging as they require a lot of effort to provide data which are results of analysis, simulations, measurements, and estimations. A complementary approach is the architectural reasoning that leads to a particular decision. For example, processing visualization data can be deliberate put on a sensor if such sensor includes an FPGA. In all cases, the process of mapping can be done in an iterative manner, in particular when fine-tuning optimization is required.


\begin{tcolorbox}
\textbf{Mapping software and hardware}


Q4: Which functionalities will be executed by the accelerator(s)?

Q5: How do the potential configurations perform?

Q6: What will be the mapping between software components and processing units?

\end{tcolorbox}

\vspace{-0.2cm}

\subsection{Determining the overall architecture design}
\label{sec:archdesign}
In this stage, the engineers analyze the requirements and constraints that were previously defined and begin to fit them into an organization that supports the system's requirements. Aspects to consider include: (i) similarities and proximity between software components; (ii) the amount of data that is transferred between components; and (iii) the use of standardized design solutions based on the type of system that is being implemented. 

There are multiple architectural design options that can be used in order to support the organization of software-intensive systems containing heterogeneous platforms. One example of such solution is the standard proposed by the HSA foundation \cite{HSA}. An HSA-compliant architecture meets the requirements for enabling heterogeneous programming models for computing platforms using standardized interfaces, processes, communication protocols, and memory models. When elaborating the overall software architecture design of such systems, \textit{communication} and \textit{computation} rise as the main aspects to be properly addressed.

C1: \textit{Address the communication aspect.}
The communication aspect is known to play an important role in heterogeneous systems since they are inherently distributed. In the case of automotive applications, there are two main characteristics that influence communication performance: the resource constraints of the embedded hardware, and the high demands on reliability that are connected to such safety critical domain. 

In AI-base systems, for instance, the processing of large amounts of data is an inherent characteristic, typically including activities to understand, structure, process and monitor information. An architecture design containing AI components must allow for the appropriate communication structures to fulfil the increasing requirements on the system. There are still few systematic methods for designing such systems, but they will be of high importance as the domain of AI advances. 

As shown in Section~\ref{sec:motivation}, computation in vehicles may be centralized, requiring access to be granted across multiple nodes. The architecture typically allows for seamless access of data, either by streaming, or on demand, in order to provide software components with the necessary means to realize functionalities. Further, there is typically a clear distinction between components that realize the training of models, and the ones realizing the execution of the models. These components must communicate, raising a number of questions, regarding e.g., the execution of these components (local, or parallel), and the re-distribution of trained models to the components that requested them.

One way to practically address the communication topic is through the \textit{separation of concerns} technique, setting ``communication’’ and ``computation’’ as main concerns in the center. Components that exchange messages are placed closer in the architecture, as well as components that are executed by the same processor. Another possibility is to establish three main concerns in the architecture: ``application model'', ``platform model'', and ``mapping between application and platform'', as presented in \cite{Andrade2012}. The application specification should contain the non-functional requirements formally encoded. The platform model should specify redundancy and replaceability of computation, as well as I/O components. The mapping should bind the application to the hardware resources according to the non-functional requirements.

Another practical possibility is to design the architecture in \textit{layers}, including a communication layer that allows different processors to communicate, as shown in \cite{Auerbach2012}. Such standardized channel of communication between different processing units allows for developers to avoid explicit handling of the low-level memory copying. Further, it is also possible to design a dedicated layer for constant monitoring of the resources, providing status information to a deployment layer.

C2: \textit{Address the computation aspect.}
The computation aspect must also be addressed, as the distribution of computational load has direct influence on the system's performance. 

As the prices of hardware components decrease, the opportunity to distribute computation between the cloud and the edge arises as an alternative to architectures based on cloud-only or edge-only computation. As mentioned earlier in Section~\ref{sec:motivation}, there is a trend in the automotive industry to move from simple CPU-based computation to smart sensors that contain powerful, heterogeneous computational nodes on the edge. The main motivation is the large amounts of data to be processed, which can be partially handled already on the edge. The important decisions in this context are related to the analysis of locality or globality of data, real-time and performance requirements, and similar concerns. For instance, certain types of data can be pre-processed already in the vehicle, while the training of models, storage of data, and execution of computationally intensive tasks can be done in the cloud.

In practice, a \textit{pipelined architecture} \cite{Gayen2007} allows the software to be represented as general data flow graphs, with particular focus on the performance. The approach bases the allocation strategy on the simulation of executing these graphs. The pipelined architecture is a reasonable candidate architectural style when there is a clear separation between the component functionality and the processing data. Part of the processing can be placed on different processing units, then the transfer of data can be defined through communication rules.

Alternatively, and most commonly, engineers can implement a \textit{master-slave architecture} in order to take advantage of the inherent characteristics of heterogeneous platforms which contain, typically, one main processor (CPU for procedural tasks), and one or more accelerators (e.g., GPU for highly parallelized and dynamic tasks). On AI-based systems, for instance, the main application flow may be processed by a CPU (master) while the training of the model is performed by the GPU, the accelerator (slave).

Further, \textit{aspect-oriented architecture}\cite{Kiczales1997} can be used in the context of building components that are executable in different processing units -- the portions in the design that are platform-specific can be treated as aspects in the overall architectural design. A typical example of realizing it is through conditional compilation, where the conditions are connected with the different processing units that are available. The approach can be used for automatic generation of code specific for a given platform, for example in creating connectors for data communication between different execution platforms \cite{Ciccozzi2013}.


\begin{tcolorbox}
\textbf{Determining the overall architecture design}


Q7: How high are the communication demands? 

Q8: How to address the distributed computation aspect?

\end{tcolorbox}

\vspace{-0.2cm}

\subsection{Refactoring software components}
Finally, the design of the individual components must be sketched and implemented according to the previously defined characteristics of the overall architecture. 

D1: \textit{Determine the new set of software components.}
In this step, engineers analyze the current architecture design and determine which components will be refactored, and which ones will be created. There might exist a number of constraints to refactoring or creating new software components due to limited hardware resources, time constraints, or increased complexity of the system. However, it is important to precisely determine the changes that will occur to every software component. Components that have migrated from one platform to another must comply with the characteristics and limitations of the target hardware architecture. Therefore, the efforts for refactoring them must be considered, as it may occur that an extensive re-design has to be performed.

D2: \textit{Design and implement the software components.}
In this step, the engineers will determine \textit{how} the software components will be either designed or refactored. This stage is crucial to architectural decisions, since the adaptation of a component to be executed on an accelerator typically requires communication structures to be created. In practice, a component that is developed for execution on CPU is likely to be turned into two components due to the nature of heterogeneous platforms. CPU remains as the host processor and executes the main flow of the application, while the most computationally intensive portion is offset to the accelerator. This scenario demands robust solutions for the communication between portions running on different processing units. As shown in the code snippet in Section~\ref{sec:motivation}, the simplest solution is to designate a shared memory accessible by both units. For complex algorithms with large amounts of data, this solution might create deficiencies in performance due to the data transfer between memory spaces from dedicated to shared.

The concept of \textit{flexible software components} can be used in order to create software components that can be executed in any of the available processing units. Support for this type of component design has been proposed earlier in the context of GPUs \cite{Campeanu2017}. Flexible software components allow developers to focus on implementing functions, while mechanisms (namely \textit{adapters}) automatically transfer data between components, taking into consideration the platform specifications. 
Creating flexible components results in higher flexibility in the architecture, allowing several execution algorithms to be implemented (e.g., round-robin, first deadline first). However, the implementation of flexible components typically includes computation overhead using adapters that is not negligible, due to the additional code transformations that are needed for execution by any processor.


\begin{tcolorbox}
\textbf{Refactoring software components}


Q9: Which will be the new component set?

Q10: How will components be refactored or created?

\end{tcolorbox}

\vspace{-0.1 cm}

The presented stages suggest a sequential decision process along the attached activities. In practice, however, the process is iterative with revisions of specific questions. For example, a decision of mapping components may lead to changes in the overall architecture, in the architectural styles, or in the type of communication (e.g., the communication between two components can be changed from serial communication over the bus to using shared memory).

\section{Validation}

%
%

\subsection{Validation procedure}
One main aspect of this framework is that we included practitioners from industrial contexts in the loop of creating the approach. Then, we conducted a set of steps in order to evaluate whether or not the proposed design approach is appropriate for its purpose, meets all constraints and will perform as expected. 

In total, we presented the framework to six companies that were in different stages in accommodating heterogeneous platforms into their processes. We presented the proposed approach and the rationale behind every step in the process. The group then discussed the initiated topic and expressed agreements and/or disagreements to every aspect that was shown. The basis for argument were typically their day-to-day activities at work and their own views on how the refactoring process should occur. After several iterations with different partners, we adjusted the framework and sent it back to the them. We also sent out a questionnaire in order to capture the respondents' background information along with their impressions of the framework. We have received written feedback from two large organizations that are market leaders in their respective industries.

The two companies we have received replies from are briefly described next. Company A is a large, globally distributed manufacturer of busses, trucks and construction equipment with strong focus on technology and innovation. It is a key player in the vehicles market and has made significant investments in the development of self-driving vehicles technology. Company B is a recent subsidiary of the automotive group that Company A is inserted in. It mainly addresses software development projects with focus on autonomous driving and driver assistant systems.

The respondents come from different backgrounds and have slightly different work assignments and experiences with heterogeneous computing. The employees of Company A are based in India and have focus on research projects related to heterogeneous computing. They are a part of mainly works on programming models to facilitate software development across different types of processors. The employee of Company B has a software development role and is currently working on computer vision algorithms, with focus on object detection. The employee reported some experience on high performance computing, although limited expertise on heterogeneous platforms. Both companies develop embedded systems in the automotive domain, and utilize GPUs for acceleration.

\subsection{Received feedback}
The questionnaire that was sent out contained three questions regarding the professionals' backgrounds and experience, followed by a general open question about the proposed framework. Then, two questions were added regarding their opinions about adding or removing aspects of the framework. Finally, there were questions about the architectural decisions that are typically made in the context of their work.

We received feedback that was complementary to the discussions which occurred during the meetings, and they are presented as follows. 

\textit{Feedback loops:} One main aspect that was reported was the need for feedback loops between the different steps, particularly during the ``software and hardware mapping'' step, in which mechanisms like ``profiling'' are necessary prior to determining the best configuration according to a given set of requirements. The changes are typically constant and iterative, allowing smaller changes to occur at each iteration. 

\textit{Continuous refactoring:} Refactoring is regularly conducted due to constantly changing requirements, despite the high complexity of the projects. Therefore, the process should include careful analysis and assessment of the current architecture prior to making changes into effect. 

\textit{Priority to software:} The projects typically put software in the center, and later on evaluate which types of hardware are needed for execution. The functionalities that entail applications are the main focus for the development of the systems.

\textit{Dependency analysis:} In Company A, the refactoring process includes an analysis of dependencies to be conducted on the components that are meant to be executed on the accelerator(s) (in this case, GPUs). Since components are typically developed for execution on CPUs, and CPUs are inherently serial in their execution method, developers must check whether there are any dependencies that prevent algorithms from running in parallel, due to the parallel execution nature of GPUs. When there are no dependencies, the component can easier be transformed into GPU-runnable code. Otherwise, when there are dependencies, the core functionality within the component must be changed in order to make it parallel. 

\textit{Refactoring procedure:} The following procedure is followed in Company A in order to refactor. The CPU-oriented program is used as baseline for performance measurement. Once the algorithm is modified into code that can be executed on an accelerator (GPU in their case), the execution time is again measured and compared with the performance of the CPU code. The changes in the execution time is then thoroughly analyzed, followed by a process to determine whether or not such deviation is acceptable. In case the trade-offs are approved, the developer proceeds to port code to the GPU.

\textit{Execution policy:} Company A has reported that their usual policy to determining the execution of software components is heavily based on profiling. Typically, the functions that take more time to execute are selected as guidelines to determine software and hardware mapping.

\section{Related work} 
As identified in a literature reviews \cite{Andrade2019,Brodtkorb2010}, there is a large amount of literature addressing heterogeneous computing. In particular, the software deployment stage is highlighted as one of the most challenging aspects of applying this technology. Several concerns and approaches were identified in \cite{Andrade2019}, primarily addressing the problems of scheduling, software quality, and software architecture pointing to the challenges in establishing a design that balances the workload between units properly. Moreover, it was mentioned through some studies the importance of a solid communication strategy, as well as the efficient management of memory spaces.

Twenty-eight studies discussing concerns of software architectures for heterogeneous computing were identified previously in \cite{Andrade2018}. These studies typically propose solutions to specific problems, rather than a holistic framework to aid in the process of migrating to heterogeneous platforms. Some of them are described next.

In \cite{Riebler2016}, the authors tackle the problem of workload distribution according to the characteristics of both the load and the processing unit. The approach identifies \textit{hotspots} in the code, and then means to generate binary code depending on the processing units that are available.  The proposed architecture contains one component (called \textit{orchestrator}) to perform resource allocations at runtime and monitor the system.

The problem of resource allocation is also addressed in \cite{Posadas2014}, which the authors propose an approach that inputs a standard UML/MARTE model and explores different allocation possibilities for software components. From a number of different models, the proposed approach generates the software infrastructure required to connect different memory spaces using communication libraries. Another example is presented in \cite{Lai2017}, in which the authors propose a GPU interface to identify race conditions through simulations.

Other papers simply present an architecture design that includes heterogeneous platforms. In \cite{Meng2012}, for instance, the authors propose an architecture design for a CPU-GPU-FPGA-based hardware platform that is used for applications in the health domain. The solution uses the pipelined architectural style, processing images from a camera feed.


\section{Conclusion \& Future work}
Heterogeneous platforms, i.e., hardware containing processing units like CPUs, GPUs and FPGAs are now reaching accessible costs, making a reasonable case for adopting such alternative. In this sense, heterogeneous computing has emerged as a viable option to satisfy increasing system requirements, such as performance, energy consumption, and time constraints. However, the process of accommodating such hardware into the system may be challenging in a number of different aspects. One of such aspects is the software architecture design, which is very likely to be adapted for the software take full advantage of the underlying hardware. 

In this paper, we proposed a framework that supports the refactoring of software systems when migrating from CPU-based projects to execution on heterogeneous platforms. Such migration poses a number of challenges to the software architecture design, in particular the allocation of resources and management of memory spaces and communication. The framework is divided into four steps that architects should follow in order to make architectural decisions that support the newly added hardware capabilities. The steps are, namely: \textit{A. ``Determining the impact on the software architecture''}; \textit{B. ``Mapping software and hardware''}; \textit{C. ``Determining the overall architecture design''}; and \textit{D. ``Refactoring software components''}. Within the refactoring process, the engineers are guided through a set of questions that allow for considerations for re-design, focusing on architectural decisions.

The research methodology conducted as follows. First, we studied the literature and identified the common approaches for software architecture when heterogeneous platforms are available. Then, we included our expert industrial partners in the loop by conducting face-to-face workshops in order to obtain their in-practice perspectives on the matter and iteratively evolve our proposed framework. Finally, we sent out questionnaires and obtained written feedback from the participants in order to improve the approach.

As future work, we will refine and extend the proposed approach to include further considerations to the migration problem, both prior to and after the re-architecting stage. We intend to provide in-depth analysis of each step in the same way that we have done for the refactoring stage presented in this work. Further, we will evaluate the technical feasibility of the complete framework in collaboration with our partners. Then, we intend to investigate the impact of business decisions on the architectural decisions connected to the refactoring of systems to accommodate heterogeneous platforms.

\section*{Acknowledgment}
This research was supported by the research projects 
``HELPING -- Heterogeneous Platform Deployment Modelling of Embedded Systems'' funded by the Swedish Research Council, and ``HoliDev -- Holistic DevOps Framework'' funded by Vinnova.

\bibliographystyle{IEEEtran}  
\bibliography{library}

\end{document}